# Dynamics and Scaling of Reynolds Shear Stress in Adverse Pressure-Gradient Flows

T.-W. Lee* and J.E. Park

*Mechanical and Aerospace Engineering, SEMTE, Arizona State University, Tempe, AZ, 85287*

**Abstract-**Using a dynamical transport analysis for the turbulence momentum, the Reynolds stress gradient can be expressed as a function of the local momentum flux and force terms. From the perspective of an observer moving at the local mean velocity, Reynolds stress gradient is seen to represent the lateral transport of streamwise momentum, balanced by the $u'^2$ transport, pressure and viscous force terms. Data from direct numerical simulations (DNS) validate this method for adverse pressure-gradient boundary layer flows at $\beta$ = 1.4 and 39 (Kitsios et al., 2017), with a good degree of consistency and agreements. Reynolds shear stress profile, in its full attributes and minor fluctuations, are replicated through the Lagrangian momentum equation. Gradient analysis also leads to scaling at the first- and second-derivative levels, for $u'^2$, $v'^2$ and $u'v'$. These findings lead to both quantitative prescription and insights as to the origin of the Reynolds shear stress structure in adverse pressure-gradient flows.

*T.-W. Lee
Mechanical and Aerospace Engineering, SEMTE
Arizona State University
Tempe, AZ 85287-6106
Email: attwl@asu.edu

# INTRODUCTION

Adverse pressure-gradient (APG) flows occur over airfoils, compressor/turbine blades and other curved surfaces such as soccer and golf balls, with profound impact on the flow pattern and drag characteristics. Since the boundary-layer structure does not scale in the usual manner as in zero-pressure gradient, this type of flow presents a unique and interesting challenge to turbulence research. Recent advances in numerics and computing power allow for large-eddy simulations (LES) or direct numerical simulations (DNS) to reveal minute details (Inoue et al., 2013; Lee, 2017; Na and Moin, 1998; Tanarro et al., 2020) at fairly high Reynolds numbers. Also, large-scale experiments (Monty et al., 2011; Schatzmann and Thomas, 2017) add to the amount of data now available for analyses. Although some of the scaling studies (Zagrola and Smits, 1997; Pozuelo et al., 2022) reveal key controlling variables, succinct definitions and insights into the underlying physical mechanisms for the turbulence structure are still sought.

Romero et al. (2022) considers "stress equations" for scaling of the turbulence profiles in adverse pressure-gradient flows, beyond the Zagarola-Smits work (1997). Collapsing the data curves is one of the goals in turbulence research, as it provides a framework to draw out the key physical variables and to visualize how the flow structure came about. Pozuelo et al. (2022) use large-eddy simulation (LES) for APG flows up to $\beta=1.4$, and find that different scaling (ZG and edge scaling) is applicable for the near-wall and outer regions. It is interesting that the near-wall peaks of $u'^2$ occur at $y^+ \sim 15$, as in zero pressure-gradient (ZPG) wall-bounded flows. Secondary (outer) peak locations are stretched outward according to the boundary layer thickness to $y/\delta^* \sim 1.4$ (Pozuelo et al., 2022). Sung and co-workers (Yoon et al., 2020; Lee and Sung, 2008) attempt attached-eddy modeling to APG flows, by accounting for the contributions from "attached" and "detached" eddies. This is based on the idea that $u'^2$ and other stress components ($v'^2$, $w'^2$) follow logarithmic scaling in the mid-layer region within some ranges of eddy hierarchies. In APG flows there are additional effects such as the boundary layer broadening and appreciable cross-stream (normal to the wall) velocity, and scaling is not as straight-forward a matter as in zero pressure-gradient (ZPG) flows (Romero et al., 2022; Maciel et al., 2018).

For practical purposes, turbulence models are used to simulate the flow field with varying degrees of agreements with DNS data (Yorke and Coleman, 2004). The models typically involve mixing length, eddy viscosity or related parameters to functionally replicate the observed turbulence structure. In that regard, they start with an assumed mechanism for the Reynolds shear stress (u'v'). For example, the models discussed in York and Coleman (2004) are based on the hypothesis that the Reynolds shear stress is relatable to the mean velocity gradient as modulated by the turbulence viscosity.

Through a completely different perspective and logic, we have applied a fundamental momentum and energy balance for a control volume moving at the local mean velocity, to arrive at a set of turbulence transport equations for the primary components ($u'^2$, $v'^2$ and u'v') of the Reynolds stress tensor (Lee, 2018; Lee, 2021a; Lee, 2021b; Lee, 2024). The inter-relationships (Eqs. 1-3, in the next section) have been amply validated in canonical turbulence geometries: free jet (Lee, 2021a), zero pressure-gradient boundary layer flows (Lee, 2018), and channel flows (Lee, 2021a; Lee, 2021b; Lee, 2024). These transport equations reveal the dynamical origins of the observed turbulence structure in these flows (Lee, 2021a; Lee, 2024), as will be discussed in the context of current results. In this work, we extend this analysis method to APG flows, with DNS data by Soria et al. (2020) who provide their data on a website (Soria et al., 2019). This is the updated version of their earlier DNS work (Kitsios et al., 2016), to "strong" APG flows with Rotta-Clauser parameter (β) up to 39. Through comparisons with DNS data and similarity considerations, we hope to demonstrate that the Reynolds shear stress in canonical and APG flows can be determined within the same theoretical framework. In addition, analysis of the flux terms suggests that scaling exists in the gradient space similar to the ZPG case (Lee, 2021a; Lee, 2024), and we present the self-similarity characteristics of the Reynolds stress components in APG flows.

## 2. A NEW FORMALISM FOR TURBULENT FLUX BALANCE

A new formulation has been developed to relate the Reynolds stress components, by considering the momentum and kinetic energy balance for a control volume moving at the local mean velocity. The result is a set of explicit expressions for the Reynolds stress components (Lee, 2021a; Lee, 2024):

u' momentum transport:

$$\frac{d\langle u'v'\rangle}{dy} = -C_{11} U \frac{d\langle u'^2\rangle}{dy} + C_{12} U \frac{d\langle v'^2\rangle}{dy} + C_{13}\left(\frac{d^2 U}{dy^2} + \frac{d^2 u'_{rms}}{dy^2}\right) \quad (1)$$

v' momentum transport:

$$\frac{d\langle v'^2\rangle}{dy} = -C_{21} U \frac{d\langle u'v'\rangle}{dy} + C_{22} U \frac{d\langle v'^2\rangle}{dy} + C_{23} \frac{d^2 v'_{rms}}{dy^2} \quad (2)$$

u'² transport:

$$\frac{d\langle u'^3\rangle}{dy} = -C_{31} \frac{1}{U}\frac{d\langle u'v'\cdot u'\rangle}{dy} + C_{32} \frac{1}{U}\frac{d\langle (v'\cdot u'v')\rangle}{dy} + C_{33} \frac{1}{U}\left(\frac{du'_{rms}}{dy}\right)^2 \quad (3)$$

The assumptions and details of the derivation can be found in Lee (2021a; 2024), and also briefly summarized in the Appendix. Eq. 1, for example, is an expression arising from the streamwise fluctuation momentum (u') balance for a control volume moving at the local mean velocity, where the Reynolds shear stress (u'v') is interpreted as the lateral (y-direction) transport of u' momentum (Lee, 2021a; Lee, 2021b; Lee, 2024). This lateral transport is balanced by the longitudinal transport (u'²) and force terms on the right-hand side (RHS) of Eq. 1. In this formulation, d/dx gradients are transformed to d/dy, following the displacement of the fluid (Lee, 2021a; Lee, 2021b; Lee, 2024). In boundary-layer flows, turbulence profiles are stretched in the transverse direction with the growth of the layer thickness. In channel flows, by introducing a displacement effect at a slightly off-set angle, the same conversion from d/dx to d/dy can be mathematically justified (Lee, 2024). Viscous shear stress due to the mean velocity gradient is still applicable in the moving coordinate frame. $C_{i3}$ constant is the kinematic viscosity, while $C_{i1}$ and $C_{i2}$ are "displacement" constants which vary as a function of the Reynolds number as discussed later.

Current formalism (Eqs. 1-3) presents a set of relationships for the Reynolds stress components, based on momentum and energy balance for a control volume moving at the local mean velocity. Eq. 1 is the x-momentum balance, wherein the Reynolds shear stress gradient (du'v'/dy) represents the net lateral momentum flux. This is balanced by the net longitudinal flux (du'²/dy) with the mean velocity U acting as the modulating factor (due to the boundary layer displacement). One focus of current work, as noted above, is to demonstrate the expandability of the current formalism to determine the Reynolds shear stress (u'v') in adverse pressure-gradient flows. There is an appreciable mean velocity component (V) in the cross-stream direction in these flows, which modifies the momentum transport, so that Eq. 1 needs to include this effect as follows:

$$\frac{d\langle u'v'\rangle}{dy} = -(C_{1u}U + C_{1v}V)\frac{d\langle u'^2\rangle}{dy} + (C_{2u}U + C_{2v}V)\frac{d\langle v'^2\rangle}{dy} + C_3\left(\frac{d^2 U}{dy^2} + \frac{d^2 u'_{rms}}{dy^2}\right)$$

(4)

This is the momentum transport equation (for a control volume moving at the local mean velocity) for the Reynolds shear stress that we assess for the adverse pressure-gradient flows in this work. The constants ($C_{ij}$'s) in Eq. 4 are listed in Table 1 for ZPG and APG flows, and discussed later in the manuscript.

**TABLE 1.**

|  | Re | β | $C_{1U}$ | $C_{1V}$ | $C_{2U}$ | $C_{2V}$ | $C_3$ | DNS Data |
|---|---|---|---|---|---|---|---|---|
| $ZPG_S$ | 1410 | 0 | 0.075 | - | -0.25 | - | 0.00057 | Spalart (1988) |
| ZPG | 4800~5280 | 0 | 0.249 | - | -0.76 | - | 0.000145 | Soria et al. (2020) |
| APG1 | 4800~5280 | 1.4 | 0.261 | 17.5 | -1.00 | 17.5 | 0.000145 | Soria et al. (2020) |
| APG2 | 22200~28800 | 39 | 1.587 | 42 | -2.75 | 5 | 0.000012 | Soria et al. (2020) |

## 3. RESULTS AND DISCUSSION

Above formulation is first validated for a zero pressure-gradient (ZPG) boundary layer flows, with V=0 in Eq. 4. The momentum balance represented in Eq. 4 can be checked by using DNS data (Soria et al., 2020; Kitsios et al., 2016). For example, numerical differentiation is applied for $u'^2, v'^2, u'v'$ from DNS data, and the resulting gradients along with U and V are input onto the terms on the right-hand side (RHS) of Eq. 4. The sum which should equal $du'v'/dy$, according to Eq. 4, can then be directly compared with the DNS data. The validation is shown in Figs. 1(a) and (b), respectively for Spalart (1988) and Soria et al. (2020) data, where in each case (different Reynolds number, 1410 and q. 4 almost exactly reproduces the du'v'/dy obtained from DNS. The negative du'v'/dy close to the wall obviously is due to the sharp negative peak in the u'v' profile, while the gradual upward slope is represented by the positive part. Zero-crossing in du'v'/dy marks the location of the negative peak. We can see that the turns and minute nuances in the inflection are quite well replicated by the current theory in Figs. 1(a) and (b), with just the flux and force terms. Most turbulence models attempt to coerce a relationship between u'v' to dU/dy or strain rates, which in hindsight has a weak observational or physical basis. Thus, much work is needed to write the turbulence viscosity as a complex function of other turbulence parameters. Instead, as in Eq. 4, it is logical and visualizable that du'v'/dy is related directly to du'$^2$/dy and dv'$^2$/dy, all momentum flux terms. A momentum balance (Eq. 4) leads to an elegant and accurate representation of the Reynolds shear stress. Similar validations for channel and other canonical flows have been extensively described in prior publications (Lee, 2021a; Lee, 2021b; Lee, 2024).

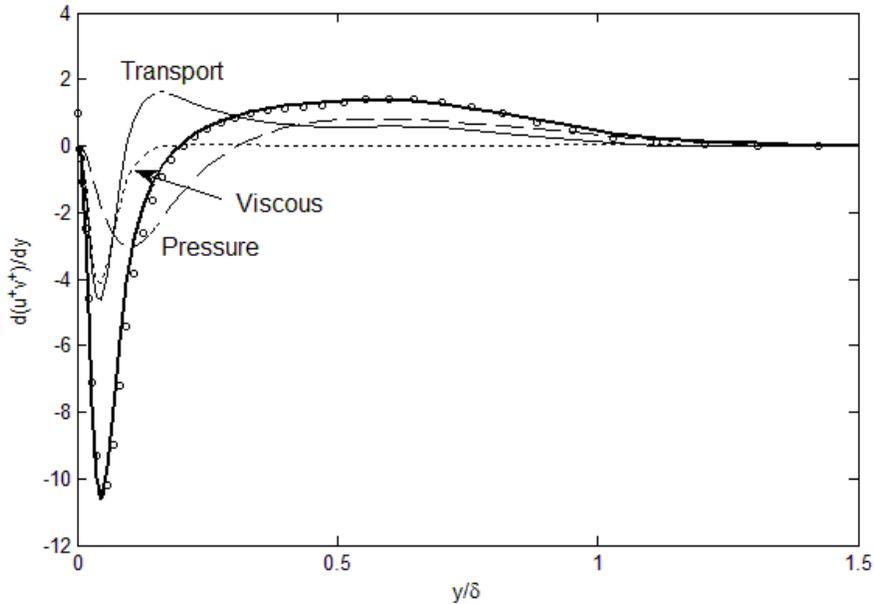

(a)

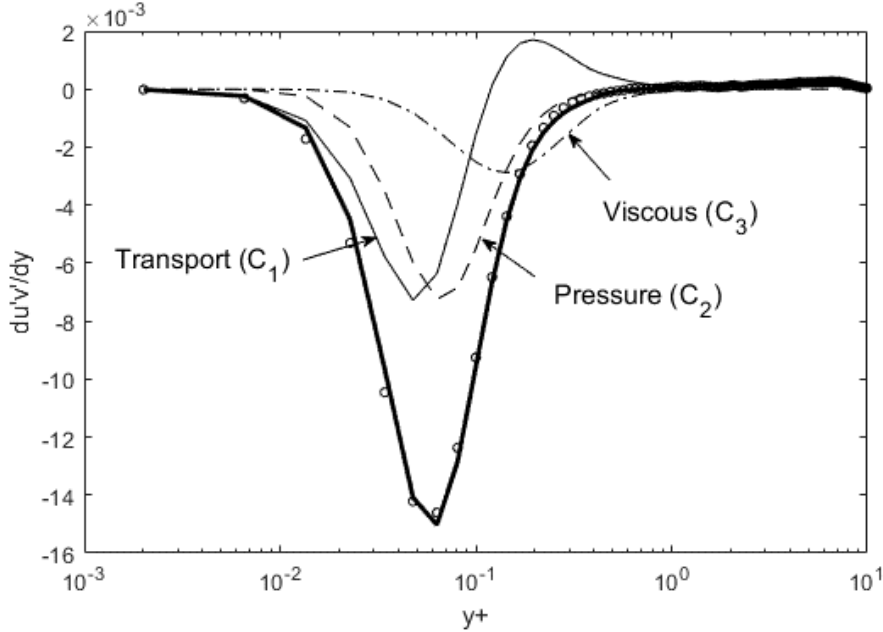

(b)

**Fig. 1. Reynolds stress gradient budget for flow over a flat plate at zero pressure gradient. DNS data (symbol) are from (a) Spalart (1998), Re=1410 and (b) Soria et al. (2020), Re=4800. Solid line is the prediction from Eq. 4. Contributions from the u'$^2$ transport ($C_{1i}$ term), pressure ($C_{2i}$), and viscous ($C_3$) terms are also plotted.**

Next, we turn our attention to the current focus: applicability of the Lagrangian transport dynamics for u'v' in APG flows. We take the DNS data by Soria et al. (2020), and again follow the same procedure of evaluating the RHS and comparing du'v'/dy directly computed from DNS. As noted in the introduction, a good theory should be applicable to other flow configurations by incorporating appropriate dynamical terms, in this case the V component. Eq. 4 represents the modified Lagrangian transport equation including this appreciable V velocity in APG flows. In addition to the streamwise displacement effect (see Appendix), there is also the relative motion of the control volume due to V that results in one additional term ($C_{1,V}$) in Eq. 4. Again, constants are tabulated in Table 1, and they depend on the Reynolds number as there is proportionately larger displacement effect.

In Figs. 2(a) and (b), current theoretical results are compared with APG flows at $\beta = 1.4$ and 39, respectively. Contributions from the individual RHS terms are also plotted to gain some insight into the momentum flux processes leading to the Reynolds shear stress structure. Agreements between theory and DNS are quite good at both APG conditions in Figs. 2(a) and (b). Large inflections and subtle nuances in the $du'v'/dy$ structure are reproduced by the theory, with only a minor misalignment for the near-wall negative peak for $\beta = 39$ at $y+ \sim 0.1$. The largest deviation occurs near y+ of 1 ~ 3 for $\beta = 39$. Overall, however, the large negative slopes and the secondary peaks overlap for DNS (symbols) and theory (line), including the fluctuations at $\beta = 39$ near y+ ~ 5 to 20. The fact that these fluctuations are replicated by the theory is another confirmation that the physical processes are properly represented in Eq. 4. Combined contributions from $C_{1U,V}$, $C_{2U,V}$ and the $C_3$ (viscous) terms in Eq. 4 clearly trace the final $du'v'/dy$ profiles in Fig. 2. The "transport" term ($C_{1U,V}$) is the main momentum flux term, which is balanced by the u'v' lateral flux ($du'v'/dy$). The pressure term ($C_{2U,V}$) consists of the $dv'^2/dy$ modified by the displacement effects from U and V, and it is responsible for forming the final shape for the near-wall negative peak and also plays the dominant role in affecting the far-wall positive peak, as shown in Fig. 2. The viscous ($C_3$) term affects the overall momentum balance and the final shape for the $du'v'/dy$, but as expected far less so at higher Reynolds number (Fig. 2b).

We can also integrate $du'v'/dy$ to compare with the Reynolds shear stress (u'v') directly (Fig. 3). There are some deviations due to numerical integration errors; a minute difference in the gradient will result in under- or over-shoot in u'v' profiles. Nonetheless, the agreement is again quite good, and the overshoot toward the free-stream boundary condition can be corrected by enforcing u'v' → 0 at large y+. Therefore, these are positive, incontrovertible confirmations of the Lagrangian transport formalism for APG flows. If we account for the turbulence momentum fluxes and force terms correctly, as in Eq. 4, then a succinct expression for the Reynolds shear stress is enabled without the need to resort to ad-hoc or complex modeling.

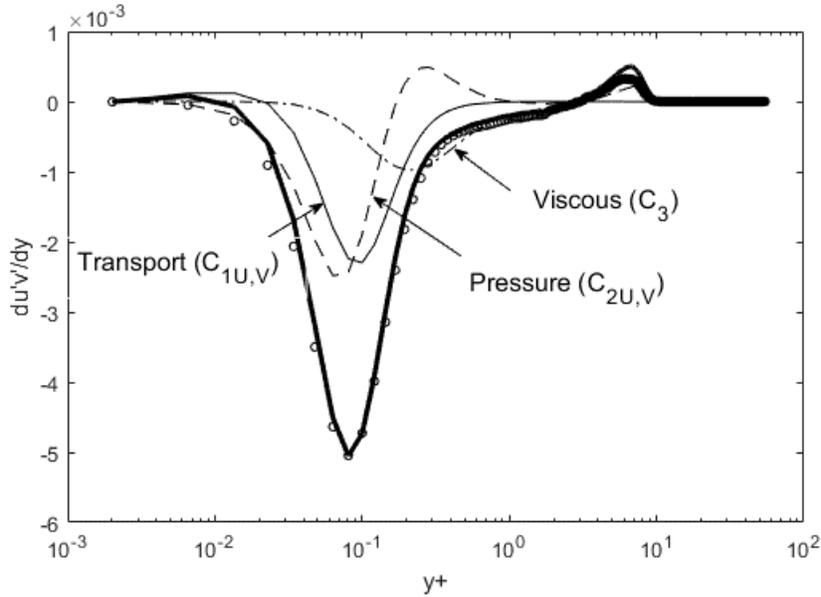

(a)

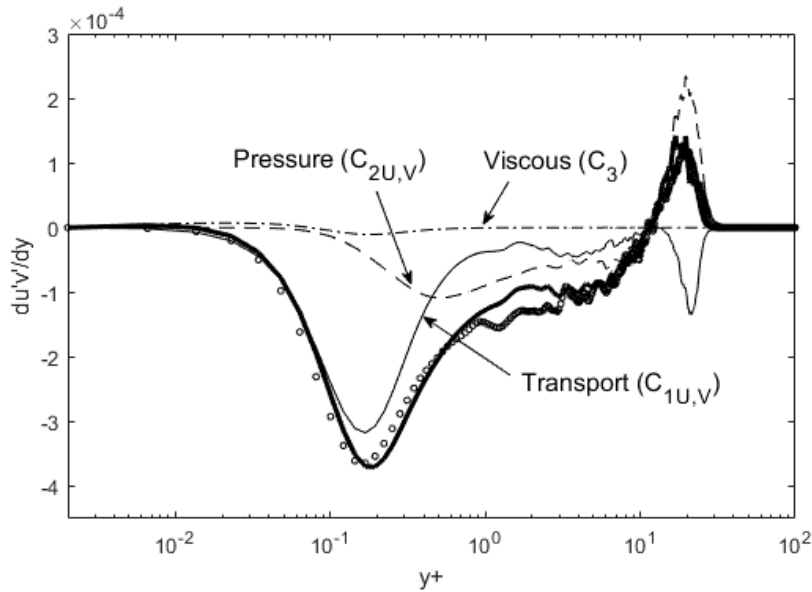

(b)

**Fig. 2. Reynolds stress gradient budget for flow over a flat plate at adverse pressure gradients, for (a) APG1 ($\beta=1.4$); and (b) APG2 ($\beta=39$). DNS data (symbol) are from Soria et al. (2020). Solid line is the prediction from Eq. 4. Contributions from the $u'^2$ transport ($C_{1U,V}$ term), pressure ($C_{2U,V}$), and viscous ($C_3$) terms are also plotted.**

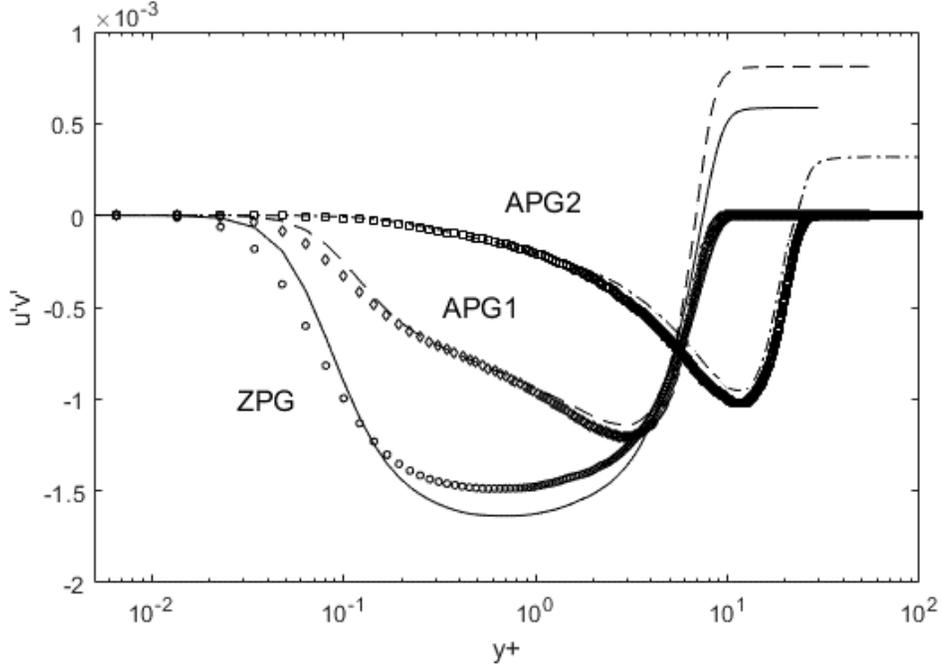

**Fig. 3. Comparison of the Reynolds shear stress for APG, ZPG1 and ZPG2. DNS data (symbol) are from Soria et al. (2020).**

The precise Reynolds number dependence for the constants in Eq. 4 (and Table 1) is a subject of a future study. In addition, the scaling, to be discussed next, may lead to just one set of constants required. For now, the constants are determined by calibrating with the DNS results, and are listed in Table 1. They increase in magnitude with increasing Reynolds numbers, indicating that the higher flow speed leads to larger displacement effects (Lee, 2021a). The constants change signs (+/-) for pressure force terms ($C_{2U}$), as they act in the opposite direction of the flux term. We can study the relative importance of the dynamical terms leading to the Reynolds shear stress through parametric variations. We double each of the terms on RHS of Eq. 4 to see their impact on the $du'v'/dy$, as shown in Fig. 4. $C_{iU}$ terms ($C_{1U}Udu'^2/dy$ and $C_{2U}Udv'^2/dy$) are doubled relative to the original $du'v'/dy$ prediction in Fig. 4(a), while $C_{iV}$ terms ($C_{1V}Vdu'^2/dy$ and $C_{2V}Vdv'^2/dy$) are doubled in Fig. 4(b). DNS data (Soria et al., 2020) and the original prediction from Eq. 4 are also included as a reference. First, doubling the $C_{1U}$ term results in significant increase (lowering) of the negative $du'v'/dy$ trace in Fig. 4(a), again illustrating the direct relationship between $du'^2/dy$ and $du'v'/dy$ for the positive/negative peaks near the wall: they have a similar structure with opposite signs. Thus, the dominant momentum balance is between $u'^2$ and $u'v'$ near the wall. $C_{2U}$ term modifies the slope in the intermediate region, between the near-wall negative peak and the far-wall secondary peak. Because the sign is reversed for the $C_{2U}$ term, doubling it results in a shallower $du'v'/dy$ profile. It is the balance between $C_{1U}$ and $C_{2U}$ terms that shapes the primary structure of the Reynolds shear stress. $C_{1V}$ and $C_{2V}$ terms both have a similar profile and contribute to the secondary positive peak (bulge), all combining to bring it to the observed (DNS) level. Without these terms, the Reynolds shear stress profiles are only qualitatively correct near the second peak; however these two terms collectively replicate the secondary peak structure.

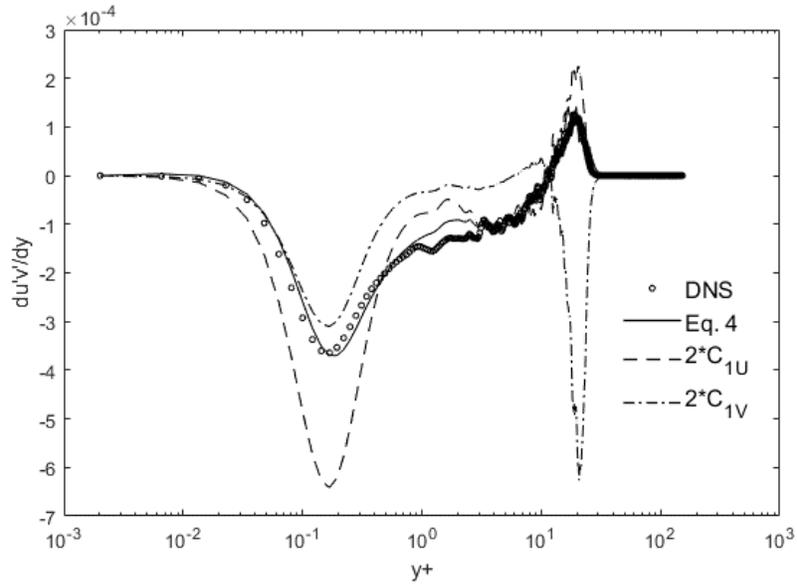

(a)

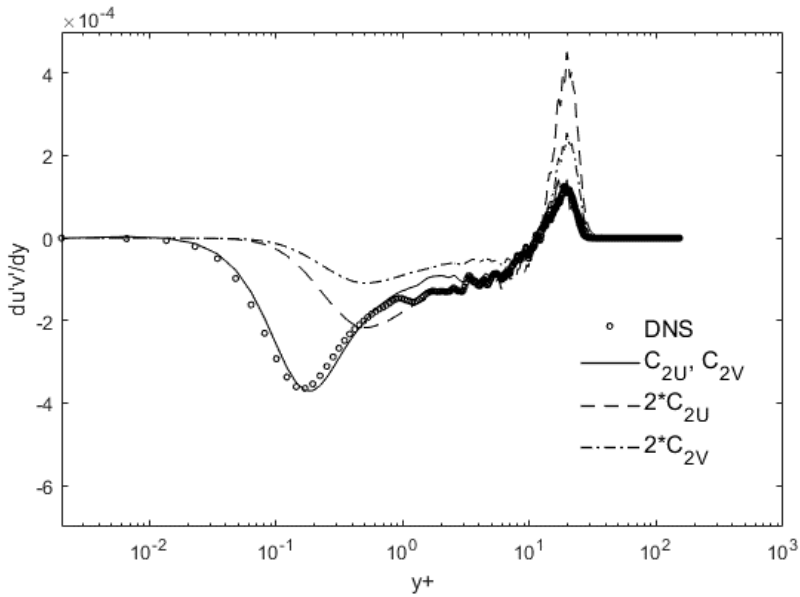

(b)

**Fig. 4. Parametric variations of the $C_{iU}$ and $C_{iV}$ in Eq. 4. DNS data are from Soria et al. (2020).**

In the preceding discussions, analyses of the momentum fluxes involved examinations of the gradient structures, which enabled reconstruction of the Reynolds shear stress profiles. This also leads to the "discovery" of self-similar characteristics in the gradient space, as shown in the next set of figures. The wall coordinates (y+) are scaled by the boundary layer thickness ($\delta$), to y+/$\delta$. Then, self-similarity features emerge for u'v' and v'$^2$, wherein full scaling is found at the second-gradient level in Figs. 5 and 6. Similar findings were made for wall-bounded flows across a large range of Reynolds numbers (Soria et al., 2020). The peak heights again scale differently for the near-wall and outer peaks in Figs. 5 and 6, suggesting that the transport processes that lead to these features are different. This is the reason the profiles are separated into near-wall (Fig. 5a and 6a) and secondary peak (Fig 5b and 6b) normalization. For the u'v' profiles that seemed difficult to pattern in Fig. 3, the full structure collapses in the second-gradient space, replicating the positive and negative concavities nearly exactly if the peak heights are normalized by the near-wall minima (Fig. 5a) and far-wall maxima (Fig. 5b), respectively. These self-similarity characteristics were first observed for wall-bounded flows (Soria et al., 2020), and it is noteworthy that the same gradient scaling extends to APG flows, which appeared much more complex on the surface.

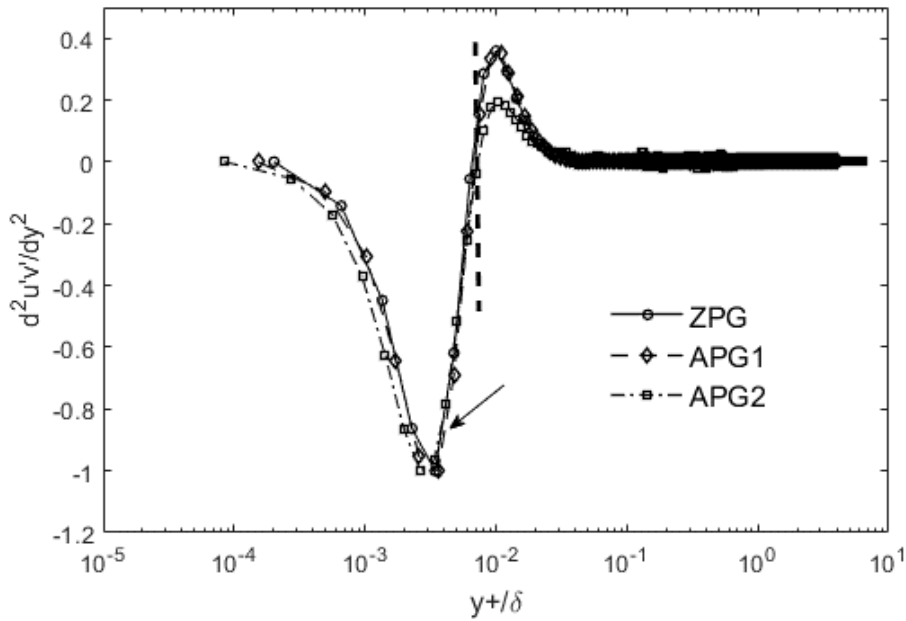

(a)

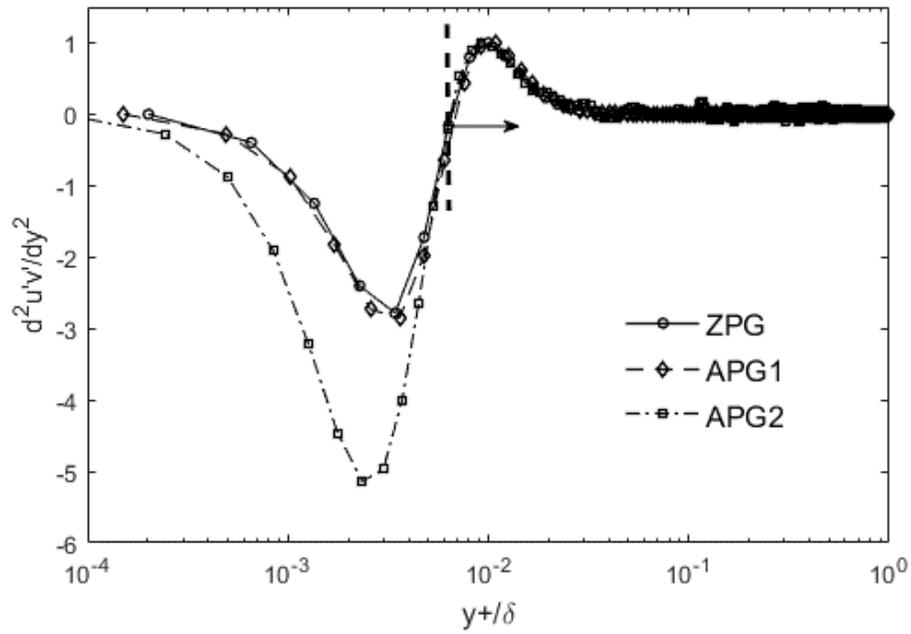

(b)

**Fig. 5. Scaling of the u'v' profiles for near-wall (a) and remaining region (b). Both the near-wall peaks (a) and far-field negative peaks (b) are self-similar at the second-gradient level, but with different normalization factors. DNS data from Soria et al. (2020) are used.**

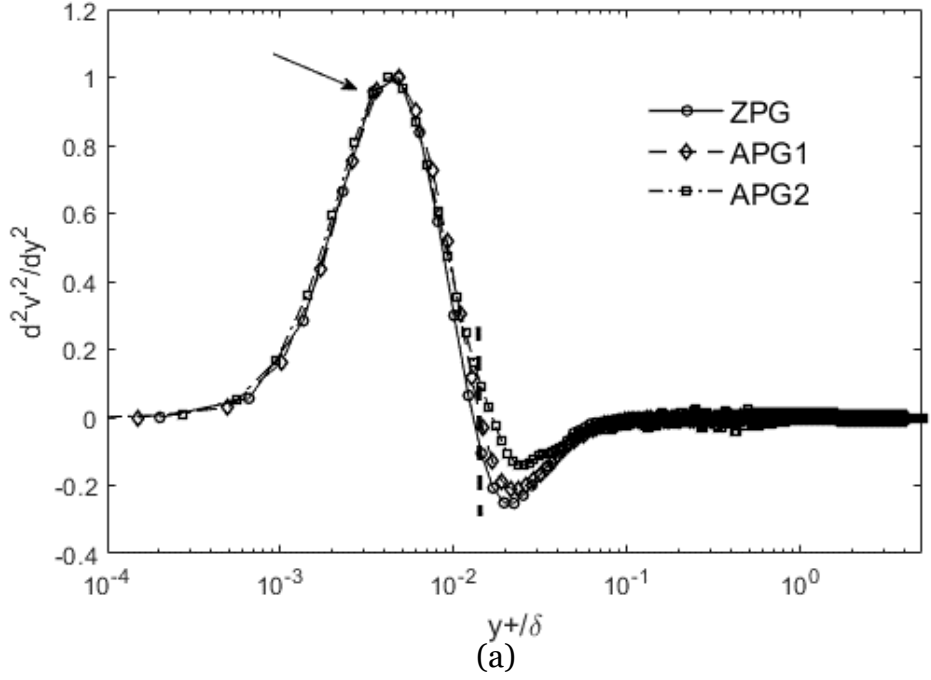

(a)

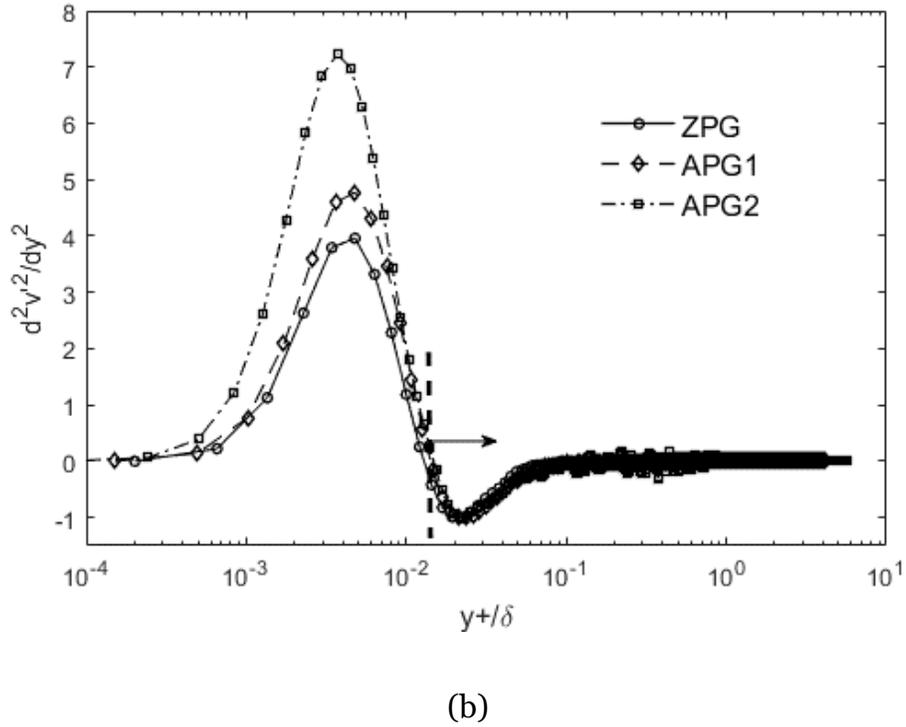

(b)

**Fig. 6. Scaling of the v'² profiles for the near-wall (a) and remaining region (b). Both the near-wall peaks (a) and far-field negative peaks (b) are self-similar at the second-gradient level, but with different normalization factors. DNS data from Soria et al. (2020) are used.**

The near-wall structure exhibits self-similarity for the *first gradient* of u'$^2$ (du'$^2$/dy) in Fig. 7(a). The outer peaks, however, scale when again the *second derivatives* are taken (Fig. 7b). Apparently, the near-wall and the outer peaks arise due to different dynamical mechanisms. Our previous hypothesis for the turbulence structure was that u'$^2$ profile is formed due to the maximum-entropy energy/dissipation distribution constrained by the wall boundary conditions (Soria et al., 2020). Then, flux and force/work terms (Eqs. 1-3, and 4) lead to spatial re-distributions. The near-wall peak location is mostly fixed at y+/δ ~ 0.003, similar to wall-bounded flows (y+ ~ 15). But the outer peak drifts slightly outward. Also, there is broadening for the near-wall peak in du'$^2$/dy at high APG. Other than those attributes, scaling is found underneath the surface in the gradient spaces for u'$^2$. Most, if not all of the scaling studies, involved working in the zeroth domain (u'$^2$, v'$^2$, or u'v' as a function of y+) and considering various statistical moments (Spalart, 1988), not in gradient spaces. Due to the structural features that arise at high Reynolds numbers or adverse pressure gradients, several additional stretching or normalization parameters were needed to align the profiles in prior studies, e.g. De Silva et al., (2015). We can see that if we start working in the gradient space, then natural progressions in the full turbulence structures emerge with only vertical (peak height normalization) and lateral stretching (by the boundary-layer thickness). The only asymmetrical feature is different maxima and minima on either side of the zero-crossing.

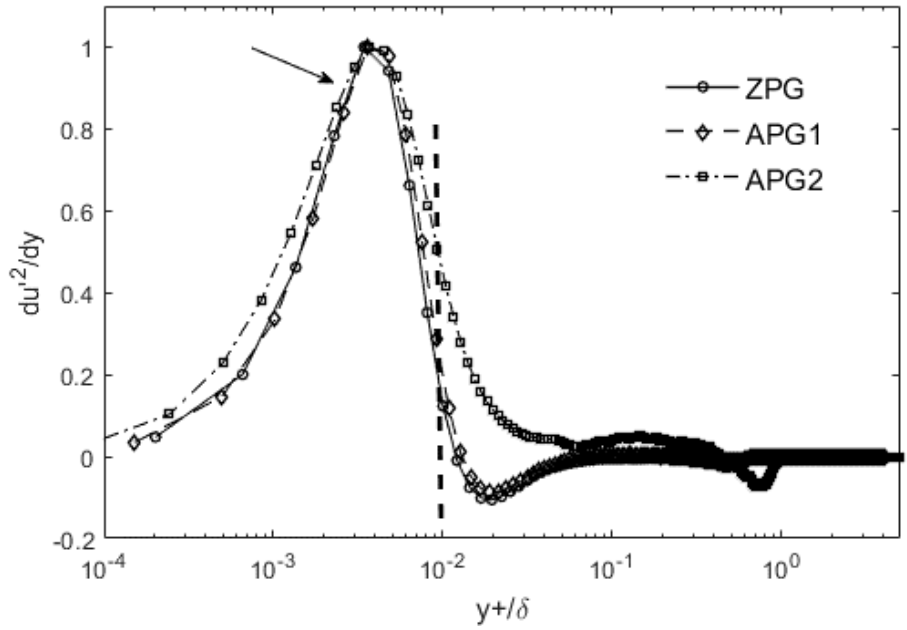

(a)

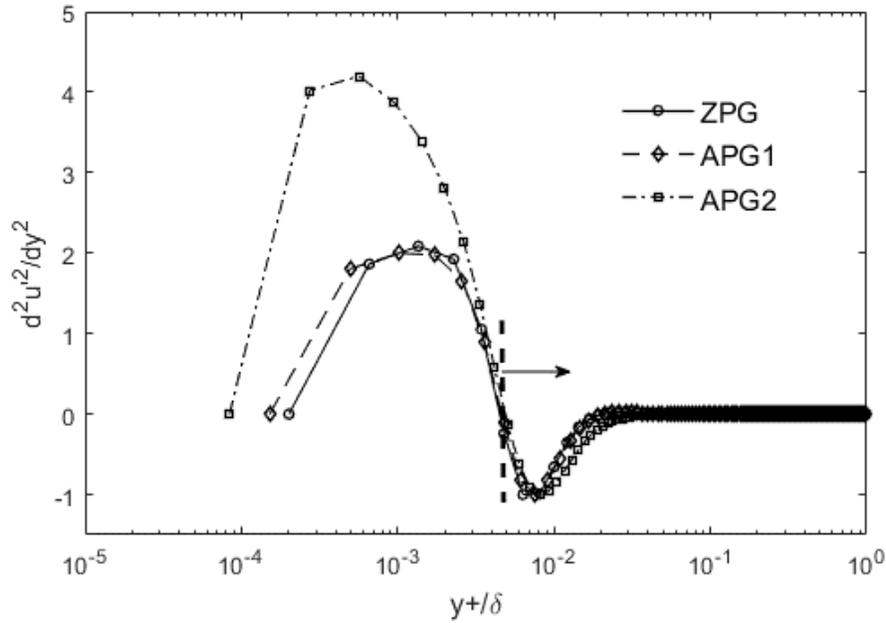

(b)

**Fig. 7. Scaling of the u'² profiles for the near-wall (a) and remaining region (b). The near-wall peaks scale as the first gradient, while the far-field negative peaks are self-similar at the second-gradient level. DNS data from Soria et al. (2020).**

**CONCLUSIONS**

The Reynolds shear stress arises due to the intricate yet rational balance between momentum fluxes and force terms, when observed from a coordinate frame moving with the local mean velocity. The most influential term is the u'$^2$ flux, or du'$^2$/dy, particularly near the wall. Other momentum components act through the pressure term, to shape the u'v' profile away from the wall. Viscous term plays a modest role in shaping the near-wall structure at low Reynolds number and $\beta$, with its importance diminishing with increasing Reynolds number, as expected. The above dynamics is based on the Lagrangian flux formalism developed for canonical flows (Lee, 2021a; Lee, 2021b; Lee, 2024), and in this work it is extended to adverse pressure-gradient boundary layer flows. Future work should test the applicability for yet more complex flows, but in its current state this formalism points toward direct dynamical relationships between the Reynolds stress gradients with fluxes and force/work terms for canonical (Lee, 2021a; Lee, 2021b; Lee, 2024) and APG flows. The origin of the Reynolds shear stress and the turbulence structure can also be seen in this manner. Full scaling of the Reynolds stress components (u'$^2$, v'$^2$, and u'v') is also observed at the second gradient level.

# APPENDIX: LAGRANGIAN MOMENTUM FLUX BALANCE

Since this formalism is relatively new, a synopsis of the logic involved is presented herein. For a control volume moving at the local mean velocity, U and V, the effects of turbulence fluctuation components (the Reynolds stress) can be isolated, as depicted in Figure 8. Eq. 1 is the transport equation for u', where the streamwise flux (u'u') is balanced by the lateral transport (u'v'), pressure and viscous forces. Figure 8(a) also shows the energy transport terms for $u'^2$, leading to Eq. 3.

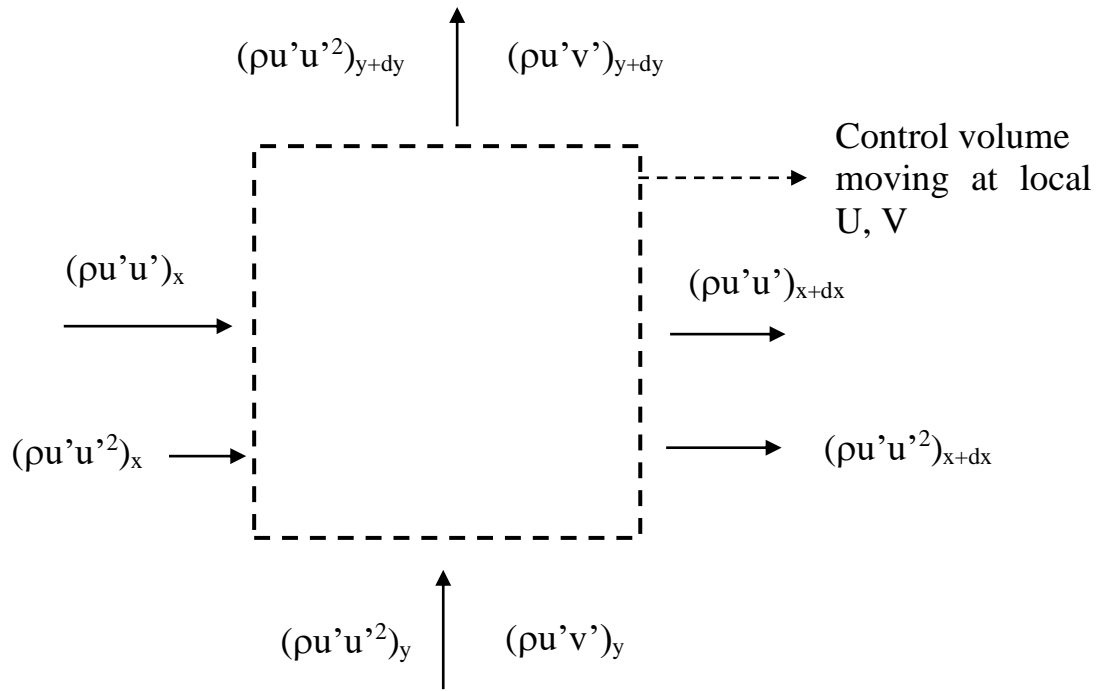

(a)

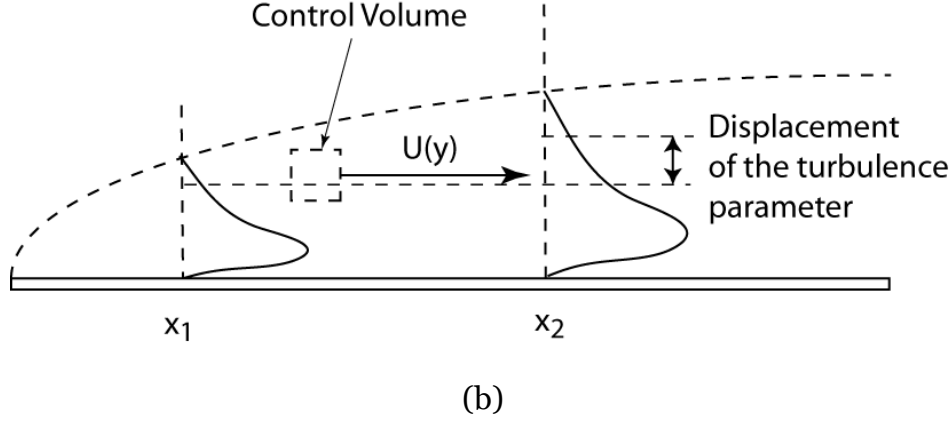

(b)

**Fig. 8. Schematic of the control volume analysis leading to the Lagrangian transport formalism: (a) x-momentum (u') and longitudinal kinetic energy balance (u'$^2$) following a control volume, moving at the local mean velocity; and (b) the displacement concept leading to d/dx → d/dy transform.**

Figure 8(b) illustrates the displacement transform used in Eqs. 1-4. Due to the displacement effect, d/dx is converted to a d/dy term with the mean velocity as the proportionality constant in this coordinate frame (Lee, 2021a; Lee, 2021b; Lee, 2024).

$$\frac{d}{dx} \rightarrow \pm C_{ij} U \frac{d}{dy} \qquad (5)$$

$C_{ij}$'s are constants with the unit, ~ $1/U_{ref}$, that prescribe this displacement effect, with U and V acting as modulating functions. The +/- sign depends on the flow geometry, i.e. the direction of displacement relative to the reference point. There are no displacement effects for channel flows since the flow is bounded by the walls on both sides. However, we still obtain the above conversion in the spatial gradients using the "probe transform" analysis. These are the new concepts leading to the transport equation set (Eqs. 1-3, and 4), and further details and validations can be obtained in the referenced articles (Lee, 2021a; Lee, 2021b; Lee, 2024) and from the authors.